\documentclass[journal]{IEEEtran}

\usepackage{cite}
\ifCLASSINFOpdf
  \usepackage[pdftex]{graphicx}
\else
\fi
\usepackage{amsmath, amssymb, bm}
\ifCLASSOPTIONcompsoc
  \usepackage[caption=false,font=normalsize,labelfont=sf,textfont=sf]{subfig}
\else
  \usepackage[caption=false,font=footnotesize]{subfig}
\fi

\newcommand*{\vv}[1]{\vec{\mkern0mu#1}}
\newcommand {\mx}[1]{_\mathrm{#1}} % normgerechte Indizierung tiefgestellt
\usepackage{siunitx}
\usepackage[hidelinks,breaklinks=true]{hyperref}
\begin{document}
%
% paper title
% Titles are generally capitalized except for words such as a, an, and, as,
% at, but, by, for, in, nor, of, on, or, the, to and up, which are usually
% not capitalized unless they are the first or last word of the title.
% Linebreaks \\ can be used within to get better formatting as desired.
% Do not put math or special symbols in the title.
\title{Driving-Cycle-Aware Shape and Topology Optimization of an Interior Permanent Magnet Synchronous Machine for a Traction Drive}

\author{Alexander Schugardt%
\thanks{The author is with the Electrical Drives Group, Technische Universität Berlin, Berlin, Germany.}}

% The paper headers
\markboth{}%
{Schugardt: Shape and Topology Optimization Framework for IPMSMs Considering Driving Cycle Performance}
% The only time the second header will appear is for the odd numbered pages
% after the title page when using the twoside option.
% 
% *** Note that you probably will NOT want to include the author's ***
% *** name in the headers of peer review papers.                   ***
% You can use \ifCLASSOPTIONpeerreview for conditional compilation here if
% you desire.

% If you want to put a publisher's ID mark on the page you can do it like
% this:
%\IEEEpubid{0000--0000/00\$00.00~\copyright~2015 IEEE}
% Remember, if you use this you must call \IEEEpubidadjcol in the second
% column for its text to clear the IEEEpubid mark.

% use for special paper notices
%\IEEEspecialpapernotice{(Invited Paper)}

% make the title area
\maketitle

\begin{abstract}
This paper presents a driving-cycle-aware shape and topology optimization workflow for interior permanent magnet synchronous machines used in traction drives. A k-means clustering approach reduces full driving cycles to representative operating points so that optimization remains computationally feasible while preserving realistic operating behavior. The workflow combines binary topology optimization, Normalized Gaussian Networks (NGnet), and spline-based shape optimization under electromagnetic, mechanical overspeed, and inverter voltage constraints. A Laplace-based mesh deformation strategy enables simultaneous optimization of magnet geometry and flux-barrier topology. Two optimized rotor designs are manufactured and tested experimentally.
The central contribution is a validated, constraint-aware optimization pipeline that achieves permanent-magnet reduction of up to 10\% while maintaining required torque capability and near-reference full-cycle efficiency.
\end{abstract}


% For peer review papers, you can put extra information on the cover
% page as needed:
% \ifCLASSOPTIONpeerreview
% \begin{center} \bfseries EDICS Category: 3-BBND \end{center}
% \fi
%
% For peerreview papers, this IEEEtran command inserts a page break and
% creates the second title. It will be ignored for other modes.
\IEEEpeerreviewmaketitle

% -----------------------------------------------------------------
% --------------------- Introduction ------------------------------
% -----------------------------------------------------------------
% The very first letter is a 2 line initial drop letter followed
% by the rest of the first word in caps.
% 
% form to use if the first word consists of a single letter:
% \IEEEPARstart{A}{demo} file is ....
% 
% form to use if you need the single drop letter followed by
% normal text (unknown if ever used by the IEEE):
% \IEEEPARstart{A}{}demo file is ....
% 
% Some journals put the first two words in caps:
% \IEEEPARstart{T}{his demo} file is ....
% 
% Here we have the typical use of a "T" for an initial drop letter
% and "HIS" in caps to complete the first word.

\section{Introduction}

\IEEEPARstart{T}{he} optimization of permanent magnet synchronous machines (PMSMs) is particularly relevant for traction drives, where rotor geometry strongly affects torque, efficiency, and power density.
Conventional parametric optimization methods restrict the design space through predefined parameters and bounds and require substantial expert knowledge to choose meaningful parameterizations~\cite{aliprantis2022,pyrhonen2013}.

Topology optimization (TO) enables novel material distributions, including new positions and shapes of air pockets, and has been applied to electromagnetic devices such as magnetic circuits and synchronous machines~\cite{lucchini2022,nishanth2022}.
Here, binary methods, boundary variation approaches, and Normalized Gaussian Networks (NGnet) are considered.

For traction drives, a full driving cycle is needed to obtain realistic efficiency and loss predictions.
This is challenging for gradient-free approaches because the large number of operating points increases computation time substantially.
Studies that combine topology/shape optimization, full-cycle evaluation, and multiphysics constraints remain comparatively scarce~\cite{cardoso2020,carraro2016,fatemi2015}.

Related work has already shown the relevance of drive-cycle-aware machine optimization and fast demagnetization assessment for traction-oriented machines~\cite{carraro2016,ferrari2023}.

This work targets a practical design objective: reducing permanent-magnet usage while enforcing traction-relevant electromagnetic, mechanical, and inverter constraints over a driving cycle.

This paper optimizes an interior permanent magnet synchronous machine (IPMSM) for an electric-scooter traction drive.
The rotor is optimized for high cycle efficiency and, in a second stage, for reduced magnet volume while respecting electromagnetic, mechanical, and inverter-related constraints.
A k-means clustering method reduces the number of operating points with minimal loss of accuracy.
The binary method, NGnet, and parameterized shape optimization are compared in terms of efficiency, magnet volume, and computational effort.
Two optimized rotor geometries are manufactured and validated experimentally using back-EMF and efficiency-map measurements.

The main contributions are: 1) a unified optimization pipeline combining cycle reduction, topology/shape optimization, and multiphysics constraint handling; 2) demonstration of up to \SI{10}{\%} magnet-volume reduction with near-reference cycle efficiency; and 3) experimental validation of the optimized designs.
In contrast to studies focused primarily on isolated operating-point gains, the present work emphasizes constraint-compliant magnet-material reduction under realistic cycle operation.

% -----------------------------------------------------------------
% -------------------------- Chapter ------------------------------
% -----------------------------------------------------------------

\section{Design requirements}

% -------------------------------------------------------------
% -------------------------------------------------------------
\subsection{Reference machine and constraints}

\begin{figure}[!t]
	\centering
	\includegraphics[width=0.48\textwidth]{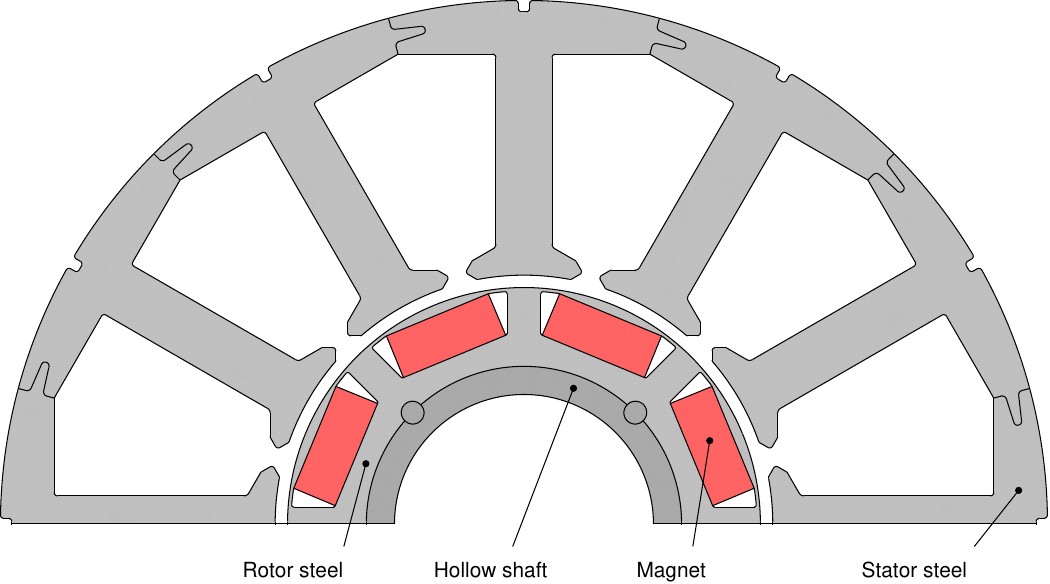}
	\caption{Geometry of the considered reference IPMSM (half cross-section)} % links, unten, rechts, oben % trim = 130 40 50 20, clip,
	\label{fig:Reference_machine}
\end{figure}

The reference machine shown in Fig.~\ref{fig:Reference_machine} is considered as the baseline for the optimization study.
It is an interior permanent magnet synchronous machine with a rated power of \SI{3.5}{kW}.
The inner part of the rotor (dark gray region) consists of a hollow shaft made of C45 steel.
The geometric dimensions, winding data, and material properties are summarized in Table~\ref{tab:machine_parameters}.
The machine is supplied by a DC-link voltage of $V_{\mathrm{DC}} = \SI{48}{V}$, and the maximum allowable phase current is limited to \SI{100}{A}.
This current limit also serves as the thermal constraint because the stator geometry and cooling system remain unchanged.
These values are used as fixed boundary conditions in all subsequent optimization steps.

%The reference machine in figure \ref{fig:Reference_machine} is considered. It is a synchronous machine with interior magnets and a nominal power of \SI{3.5}{kW}. The inner part of the rotor (dark grey) is a hollow shaft with the material C45. The geometric parameters, winding information and materials are listed in table \ref{tab:machine_parameters}. A $V\mx{DC} = \SI{48}{V}$ DC voltage is used and the current limit is \SI{100}{A}. The \SI{0.2}{\%} yield strength $R\mx{\SI{0.2}{\%}}$ of the electrical steel (M330-35A) is \SI{345}{MPa} and of C45 is \SI{490}{MPa}.

\begin{table}[!t]
	% increase table row spacing, adjust to taste
	\renewcommand{\arraystretch}{1.3}
	\caption{Main geometrical, electrical, and material parameters of the reference machine}
	\label{tab:machine_parameters}
	\centering
	\begin{tabular}{|c|c||c|c|}
		\hline
		Parameter       & Value        & Parameter                            & Value              \\
		\hline
		Axial length    & \SI{120}{mm} & Number of turns                      & \SI{31}{} per coil \\
		\hline
		Stator diameter & \SI{93}{mm}  & Parallel winding paths               & \SI{4}{}           \\
		\hline
		Rotor diameter  & \SI{42}{mm}  & Phase resistance (\SI{21}{^\circ C}) & \SI{12.9}{m\Omega} \\
		\hline
		Magnet width    & \SI{9.8}{mm} & Electrical steel                     & M330-35A           \\
		\hline
		Magnet height   & \SI{4}{mm}   & Magnet type                          & N45SH              \\
		\hline
	\end{tabular}
\end{table}

\begin{figure}[!t]
	\centering
	\includegraphics[width=0.4\textwidth]{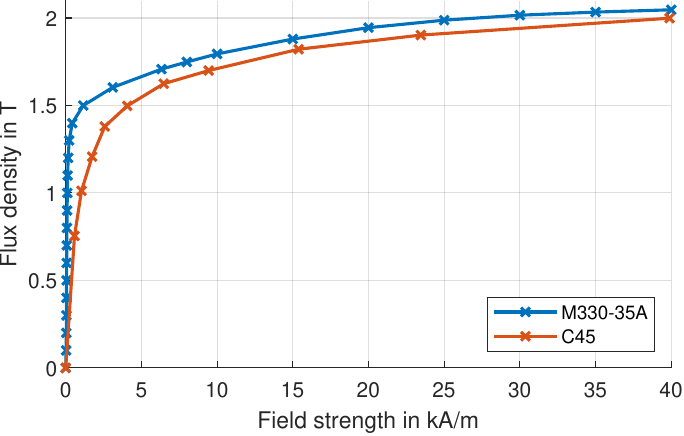} % links, unten, rechts, oben
	\caption{Measured $B(H)$-curves of M330-35A and C45 used in the electromagnetic model}
	\label{fig:BH_C45E_M330_35A}
\end{figure}

The \SI{0.2}{\%} proof strength $R_{\SI{0.2}{\%}}$ of the employed electrical steel M330-35A is \SI{345}{MPa}, while that of the C45 shaft material is \SI{490}{MPa}.
The nonlinear $B(H)$-curves of both materials are shown in Fig.~\ref{fig:BH_C45E_M330_35A} and are used in the finite-element model to account for magnetic saturation effects.

% -------------------------------------------------------------
% -------------------------------------------------------------
\subsection{Driving cycle and vehicle model}

\begin{figure}[!t]
	\centering
	\includegraphics[width=0.39\textwidth]{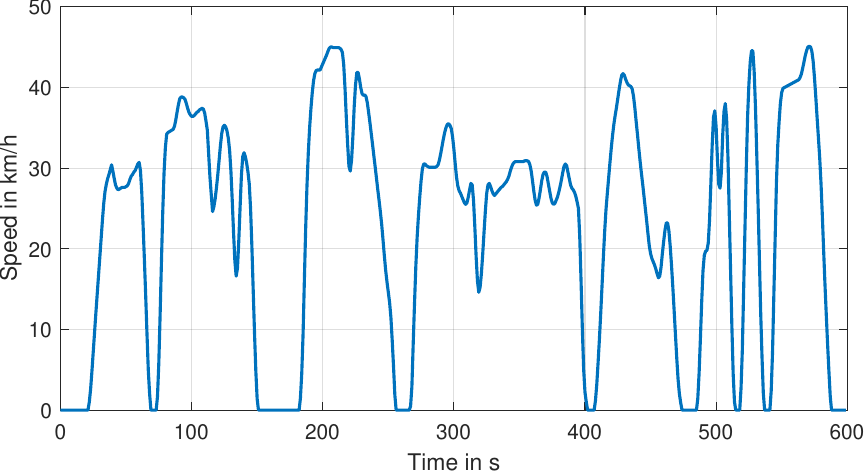}
	\caption{Considered scaled WMTC driving cycle for the electric scooter application} % links, unten, rechts, oben % trim = 130 40 50 20, clip,
	\label{fig:Driving_cycle}
\end{figure}

The machine is optimized for application as a traction motor in a small electric scooter with a maximum vehicle speed of \SI{45}{km/h}.
To reflect real operating conditions, the scaled World Motorcycle Test Cycle (WMTC) shown in Fig.~\ref{fig:Driving_cycle} is used in the optimization process~\cite{barlow2009}.
The corresponding vehicle parameters are summarized in Table~\ref{tab:Vehicle_parameters}.
A constant gear efficiency is assumed, and the longitudinal vehicle model includes aerodynamic drag, rolling resistance, and drivetrain inertia.

%The machine should be optimized for usage as a traction motor inside a small electric scooter with a maximum velocity of \SI{45}{km/h}. Therefore, the scaled WMTC driving cycle must be considered (see fig. \ref{fig:Driving_cycle}). All needed data for the vehicle model can be found in table \ref{tab:Vehicle_parameters}. A constant gear efficiency is assumed for this study. 

\begin{table}[!t]
	% increase table row spacing, adjust to taste
	\renewcommand{\arraystretch}{1.3}
	\caption{Parameters of the longitudinal vehicle model for the electric scooter}
	\label{tab:Vehicle_parameters}
	\centering
	\begin{tabular}{|c|c||c|c|}
		\hline
		Parameter          & Value               & Parameter       & Value        \\
		\hline
		Drag coefficient   & \SI{0.7}{}          & Total mass      & \SI{170}{kg} \\
		\hline
		Frontal area       & \SI{0.72}{m^2}      & Wheel radius    & \SI{0.47}{m} \\
		\hline
		Rolling resistance & \SI{0.015}{}        & Gear ratio      & \SI{13.07}{} \\
		\hline
		Moment of inertia  & \SI{0.56}{kg \ m^2} & Gear efficiency & \SI{97}{\%}  \\
		\hline
	\end{tabular}
\end{table}

\section{Framework}

%\lipsum[15]

% -------------------------------------------------------------
% -------------------------------------------------------------
%\subsection*{Algorithm}

The optimization framework uses a gradient-free single-objective genetic algorithm (GA) with a weighted sum of normalized objective terms~\cite{petrowski2017}.
Each run starts from a predefined initial geometry; for all methods except the shape-based approach, this geometry contains no air pockets.
Design parameter bounds are set to ensure manufacturability and compliance with the available installation space.

Material and air regions are assigned according to the selected optimization method, after which the machine model is generated and checked for intersecting edges and invalid domains.
Valid geometries undergo magnetostatic simulations at multiple rotor positions and mechanical stress analysis at maximum operating speed increased by \SI{20}{\%}.
The transient demagnetization check is applied only as a post-optimization verification step because of its high computational cost.

The optimization loop generates new candidate geometries using crossover and mutation operators.
It terminates after 15 consecutive generations without improvement or when a maximum generation count is reached.
A population size of 200 individuals is used for all optimization runs, and one complete optimization run requires approximately four days of computation time.

%The optimization is based on a gradient free stochastic approach. In this study, a genetic algorithm is used for this purpose. A start geometry is considered at the beginning of every optimization. No air pockets are inside this geometry for all approaches except the shape optimization. Also, parameter limits are defined for the magnet optimization or the shape approach. Material and air are set inside the first step based on the chosen method. The model is build and checked for geometry problems like intersecting lines. The geometry is ignored if any errors are found. The following step involves the magnetostatic simulation for different rotor angles and the mechanical simulation for the maximum speed (+ \SI{20}{\%}). All solutions are used to calculate a value based on the objective function. The process repeats starting with building a new geometry. If no better geometries are found over 15 successive generations of the genetic algorithm the process stops.

\begin{figure}[!t]
	\centering
	\includegraphics[trim = 0 0 600 0,  clip, clip,width=0.34\textwidth]{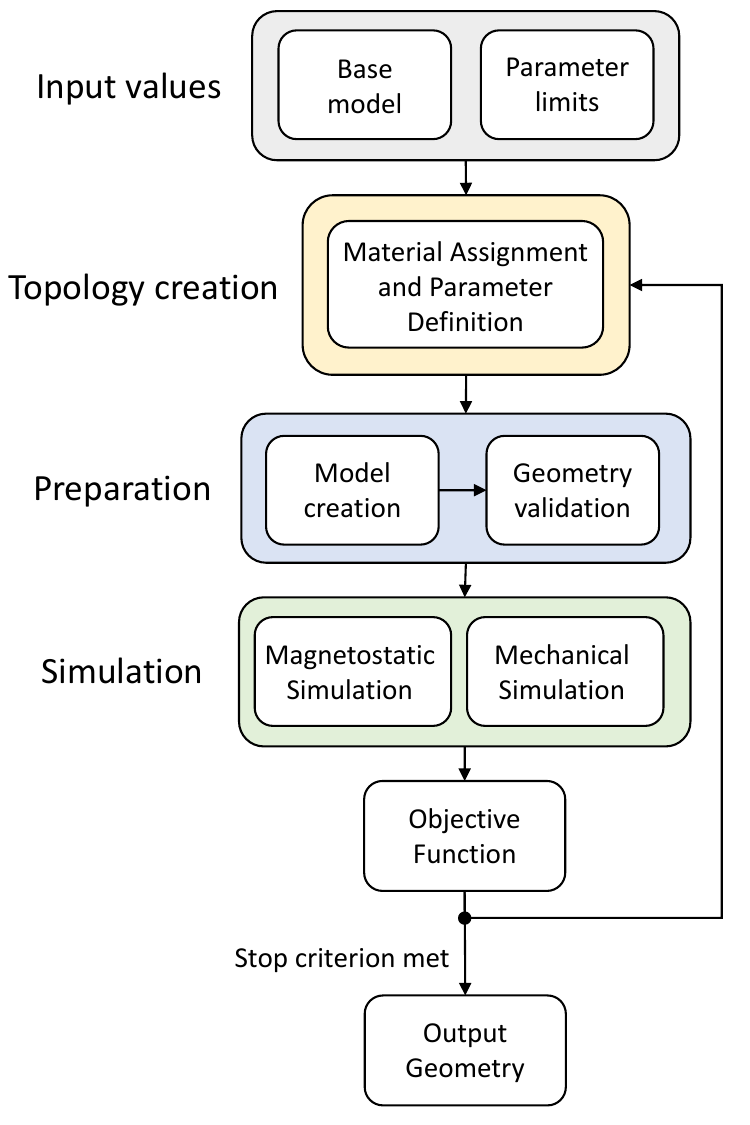} % links, unten, rechts, oben
	\caption{Proposed optimization framework for shape and topology optimization of the rotor}
	\label{Ablaufdiagramm_Optimierung_ges}
\end{figure}

% -------------------------------------------------------------
% -------------------------------------------------------------
\subsection{Electromagnetic simulation and loss model}

Torque and core losses are evaluated using two-dimensional magnetostatic finite-element simulations at multiple rotor positions per operating point.
The operating points use maximum torque per ampere (MTPA) current references; candidates that violate the voltage limit are penalized in the objective function.
Iron losses are computed with the Bertotti model~\cite{bertotti1983,krings2010}, and magnet eddy-current losses are approximated in two dimensions while three-dimensional end effects are neglected.

% -------------------------------------------------------------
% -------------------------------------------------------------
\subsection{Demagnetization calculation}

The transient short circuit is considered for the demagnetization calculation. Therefore, the voltage equations in the dq-system are discretized. The phase resistance $R\mx{ph}$, electrical rotational frequency $\omega$, number of pole pairs $p$ and time steps $\Delta t$ are used to calculate the flux linkages for the next step $k+1$:
\begin{align}
	\Psi\mx{d}^{k+1} & = \Psi\mx{d}^{k} + (-R\mx{ph} \cdot i\mx{d}^k + p \cdot \omega \cdot \Psi\mx{q}^k) \cdot \Delta t \\
	\Psi\mx{q}^{k+1} & = \Psi\mx{q}^{k} + (-R\mx{ph} \cdot i\mx{q}^k - p \cdot \omega \cdot \Psi\mx{d}^k) \cdot \Delta t
\end{align}

Inverted flux maps are used to calculate transient currents in the dq-systems. The operating point with maximum power is considered as a starting point. The highest negative d-current of the transient curve is identified and used for the demagnetization calculation. A comparison of the lowest flux density value inside the magnets with the knee flux density of the magnet material at \SI{120}{^\circ C} is conducted to verify that partial demagnetization does not occur under worst-case short-circuit conditions~\cite{ferrari2023}. Because this check is applied only after the GA search, demagnetization robustness is verified for final candidates but is not used as a hard constraint during every generation.

% -------------------------------------------------------------
% -------------------------------------------------------------
%\subsection*{Mechanical simulation}
%
%The mechanical simulation is done by solving the differential equation
%\begin{align}
%	 -\nabla \cdot \bm{\sigma} = \vv{f}
%\end{align}
%in two dimensions (plain stress) for the node displacements. The stress tensor $\bm{\sigma}$ and external forces $\vv{f}$ are used. 

% -------------------------------------------------------------
% -------------------------------------------------------------
\subsection{Objective function}

The primary optimization objective is the maximization of machine efficiency, which is equivalent to the minimization of the total losses $P_{\mathrm{loss}}$.
The secondary objective is the minimization of the permanent magnet volume, expressed by the reduction of the axial cross-sectional area of the magnets $A_{\mathrm{mag}}$.
Reference values for both objectives are obtained from the reference machine evaluated over the full driving cycle.

The overall goal of the optimization is the minimization of the combined objective function $Z$. The definitions of the first two objective function terms are given as follows:
%The first objective is the maximization of the efficiency, which corresponds to the minimzation of the losses $P\mx{loss}$. The second one is the minimization of the magnet volume, which is equal to the minimization of axial cross-section area of the magnets $A\mx{mag}$. The reference values are calculated using the reference machine. The minimization of the full objective function $Z$ is the optimization goal. Definitions of the first two summands are given as: 
\begin{align}
	%\text{Drehmoment:} \quad Z\mx{M} &= -\frac{M - M\mx{ref}}{\SI{0.2}{} \cdot M\mx{ref}} + 1 \\[5pt]
	\text{Losses:} \quad Z\mx{loss}     & =  \frac{\frac{P\mx{loss}}{P\mx{loss,ref}} - 0.8}{0.4} \\[5pt]
	\text{Magnet area:} \quad Z\mx{mag} & =  \frac{\frac{A\mx{mag}}{A\mx{mag,ref}} - 0.8}{0.4}
\end{align}
The normalization maps a value of 0.8 times the reference value to zero and a value of 1.2 times the reference value to one, while the reference design maps to 0.5.
The voltage limit is formulated using the admissible fundamental phase-voltage amplitude $\hat{v}\mx{1,lim}=m\mx{a,max}\cdot V\mx{DC}/2$.
%The scaling leads to a value of zero for \SI{0.8}{} times the reference value or one for \SI{1.2}{} times the reference value. The constraint summands for the objective function are defined based on the corresponding limits:
\begin{align}
	Z\mx{T}    & =
	\begin{cases}
		1000 \cdot \left( \frac{T\mx{ref}}{T} - 1 \right), & T < T\mx{ref} \\
		0, & \text{otherwise}
	\end{cases}
	\\[5pt]
	Z\mx{V}    & =
	\begin{cases}
		1000 \cdot \left( \frac{\hat{v}\mx{1,max}}{\hat{v}\mx{1,lim}} - 1 \right), & \hat{v}\mx{1,max} > \hat{v}\mx{1,lim} \\
		0, & \text{otherwise}
	\end{cases}
	\\[5pt]
	Z\mx{mech} & =
	\begin{cases}
		1000 \cdot \left( \frac{\sigma\mx{m,max}}{R\mx{\SI{0.2}{\%}}} - 1 \right), & \sigma\mx{m,max} > R\mx{\SI{0.2}{\%}} \\
		0, & \text{otherwise}
	\end{cases}
\end{align}

The torque, voltage, and mechanical-strength constraints are implemented as penalty terms with a factor of \SI{1000}{} whenever a limit is violated.
The soft objective terms remain of order unity, while infeasible designs are ranked behind feasible ones.

The constraint evaluations are based on the PWM modulation index $m\mx{a}$, the maximum fundamental voltage amplitude $\hat{v}\mx{1,max}$, and the maximum von Mises stress $\sigma\mx{m,max}$.
The maximum allowable modulation index is set to $m\mx{a,max} = 2/\sqrt{3}$ for space vector modulation.

The complete objective function is defined by incorporating the weighting factor $k_{\mathrm{w}}$ for the two primary optimization goals as follows:
%At first the torque limit is included using $Z\mx{T}$. If the design does not reach the needed torque for the operating point with the highest torque value, a factor of \SI{1000}{} is used as penalization inside the objective function. Similar definitions are used for the voltage limit $Z\mx{V}$ and the mechanical strength limit $Z\mx{mech}$. The PWM modulation index $m\mx{a}$, the voltage amplitude $\hat{v}\mx{1,max}$ and the maximum von Mises stress $\sigma\mx{mises,max}$ are used. The full objective function is defined under consideration of the weighting $k\mx{w}$ for the two optimization goals:
\begin{align}
	Z & = k\mx{w} \cdot Z\mx{loss} + (1-k\mx{w}) \cdot Z\mx{mag} \nonumber\\
	  & \quad + Z\mx{T} + Z\mx{V} + Z\mx{mech}
\end{align}
A weighting factor of $k\mx{w} = 0.3$ is applied in all optimization runs.
This value is selected as an engineering compromise to prioritize permanent-magnet reduction while keeping cycle losses close to the reference level; it should not be interpreted as a globally optimal weight for other applications.
When the magnet geometry is fixed, the term $Z\mx{mag}$ is identical for all candidate designs and therefore has no influence on the ranking; in this case the loss objective $Z\mx{loss}$ effectively drives the optimization.

% -------------------------------------------------------------
% -------------------------------------------------------------
\subsection{Filtering and smoothing}

\begin{figure}[!t]
	\centering
	\includegraphics[trim = 0 20 390 0, clip,width=0.48\textwidth]{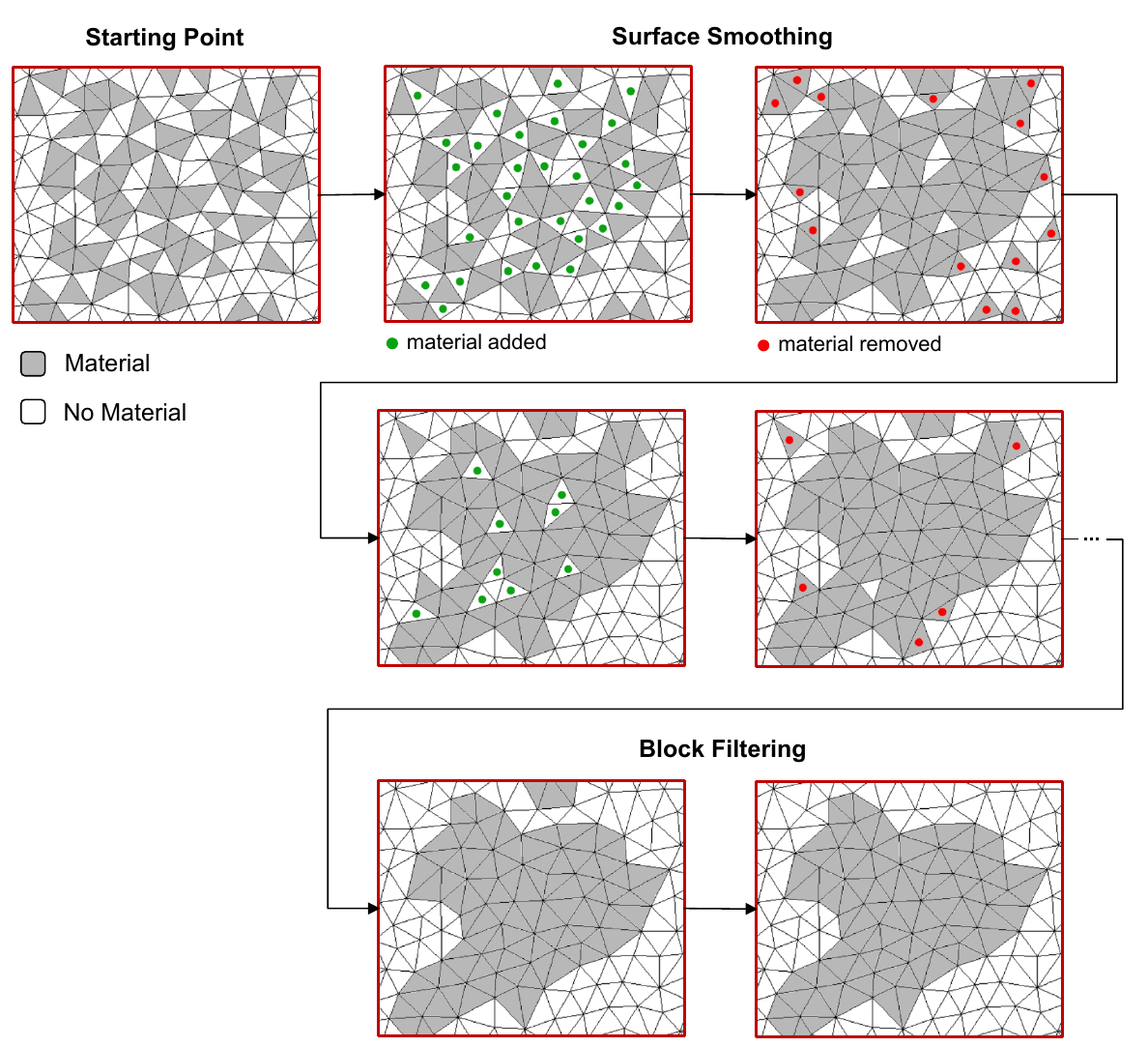} % links, unten, rechts, oben
	\caption{Visualization of surface smoothing and block filtering used as part of the gradient-free topology optimization (based on \cite{sun2023improved,sun2023hybrid})}
	\label{fig:Filterung_Dreiecke}
\end{figure}

The binary optimization method can generate checkerboard patterns or disconnected material regions, so a two-stage filtering scheme is applied.
The first stage performs surface smoothing, while the second stage implements block filtering.
An example is shown in Fig.~\ref{fig:Filterung_Dreiecke}.

During surface smoothing, all elements (triangles) with two or more neighboring elements of the opposite material are modified to reduce abrupt transitions.
The block filter removes all connected material regions except the largest one.
%Especially the binary method leads to many geometries that are not suitable for a rotor design, because it can generate checkerboard structures or multiple geometric parts that are not connected. Therefore, a filtering scheme is applied. The first step is a surface smoothing and the second a block filtering. Both are visualized in figure \ref{fig:Filterung_Dreiecke} for a random geometry. The surface smoothing influences all elements (triangles) with two or more neighbors with the opposite material. The block filter removes all connected material areas except the largest.

%\begin{figure*}[!t]
%	\centering
%	\includegraphics[trim = 40 100 50 150, clip,width=0.8\textwidth]{figures/Filterung_Dreiecke_(engl).pdf} % links, unten, rechts, oben
%	\caption{Visualization of surface smoothing and block filtering used as part of gradient-free topology optimization (based on \cite{35} and \cite{73})}
%	\label{fig:Filterung_Dreiecke}
%\end{figure*}

\begin{figure}[!t]
	\centering
	\includegraphics[trim =  10 120 70 130,  clip, width=0.48\textwidth]{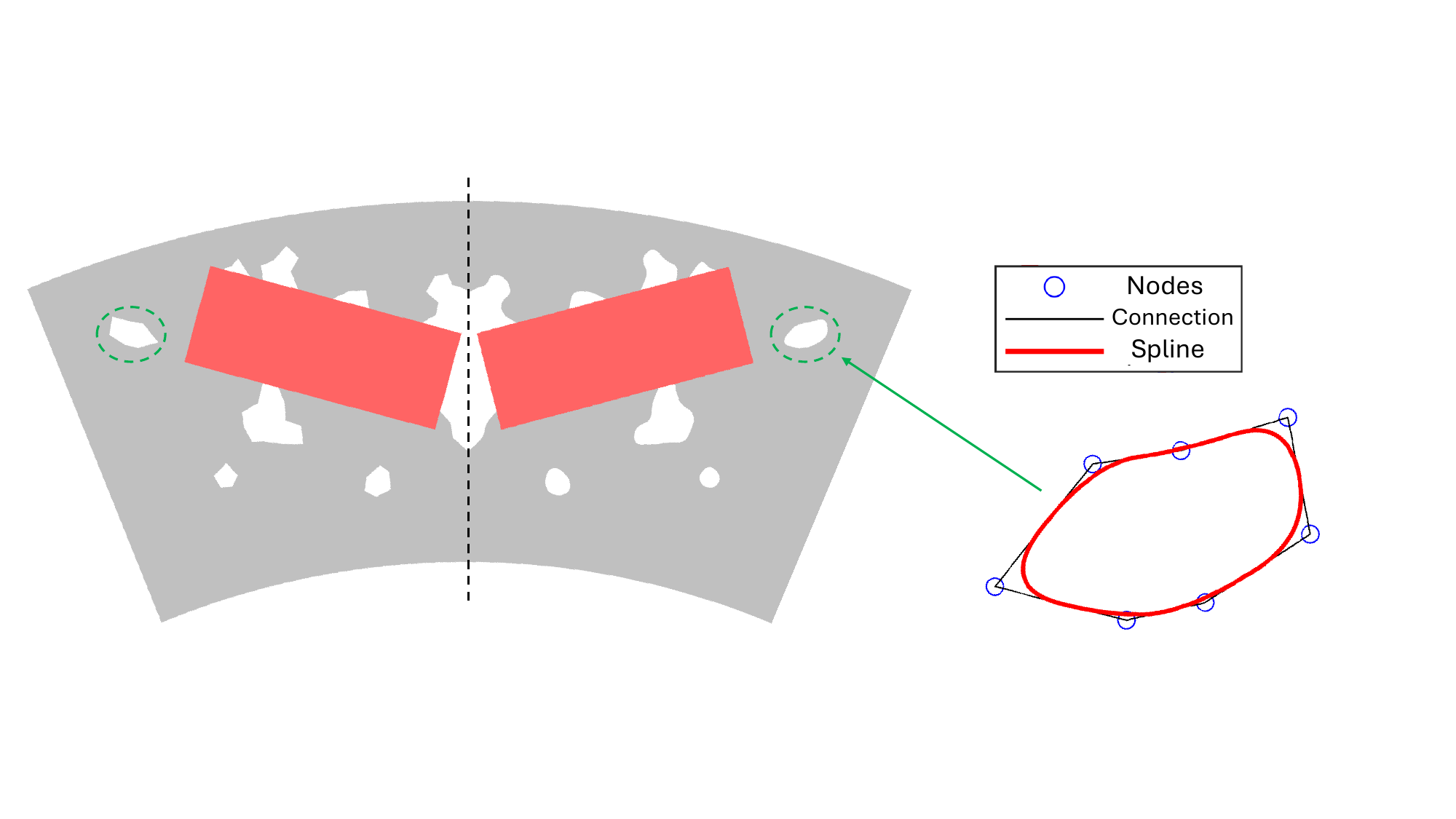}
	\caption{Spline-based smoothing of air-pocket boundaries obtained from the binary topology optimization} % links, unten, rechts, oben % trim = 130 40 50 20, clip,
	\label{fig:Lufttasche_Smoothing_Vgl}
\end{figure}

Despite filtering, the geometries still exhibit sharp edges due to the triangular mesh.
An air-pocket smoothing method based on spline interpolation is therefore introduced, as shown in Fig.~\ref{fig:Lufttasche_Smoothing_Vgl}, to reduce local stress concentrations and improve manufacturability.

%Despite the use of the demonstrated filtering, the resulting geometry still has edges because of the used triangular elements. An air pocket smoothing is introduced based on spline smoothing. The effect on the material edges can be seen in figure \ref{fig:Lufttasche_Smoothing_Vgl}.

% -------------------------------------------------------------
% -------------------------------------------------------------
\subsection{Mesh deformation} \label{ch:Mesh_deform}

In the binary optimization method, the mesh elements within the design region are treated as design variables. Since interior magnets are used, the magnet is included within the design region. If the magnet's dimensions or position are modified, a new mesh must be generated, which changes both the number of elements and their spatial arrangement.

To address this limitation, a mesh deformation strategy is employed. Using Laplace-based smoothing, the new node coordinates $(x,y)$ are computed from the initial coordinates $(X,Y)$ by solving the following two differential equations~\cite{lamecki2016,yang2016mesh}:
%The binary method handles mesh elements inside the design region as variables. The magnet lies inside the design region, because interior magnets are used. If the dimensions or position of the magnet is changed, a new mesh is created. So the number of elements, the numbeing and their relative positioning can change. A stochastic algorithm cannot be used in this case. To overcome this limitation, a mesh deformation strategy is implemented. Based on laplace smoothing, the new node coordinates $(x,y)$ are calculated based on the initial coordinates $(X,Y)$ by solving the two differential equations:
\begin{align}
	\frac{\partial^2 x}{\partial X^2} + \frac{\partial^2 x}{\partial Y^2} = 0 \\[10pt]
	\frac{\partial^2 y}{\partial X^2} + \frac{\partial^2 y}{\partial Y^2} = 0
\end{align}

The finite-element method is employed to solve these equations. Nodes that are not intended to deform are enforced as Dirichlet boundary conditions, and the magnet nodes are treated with prescribed rigid displacements.

For large magnet displacements, some triangular elements may overlap. To prevent this, auxiliary lines are introduced from fixed nodes to edge nodes of the magnets.

\begin{figure}[!t]
	\centering
	\includegraphics[trim = 0 30 50 0,  clip, width=0.48\textwidth]{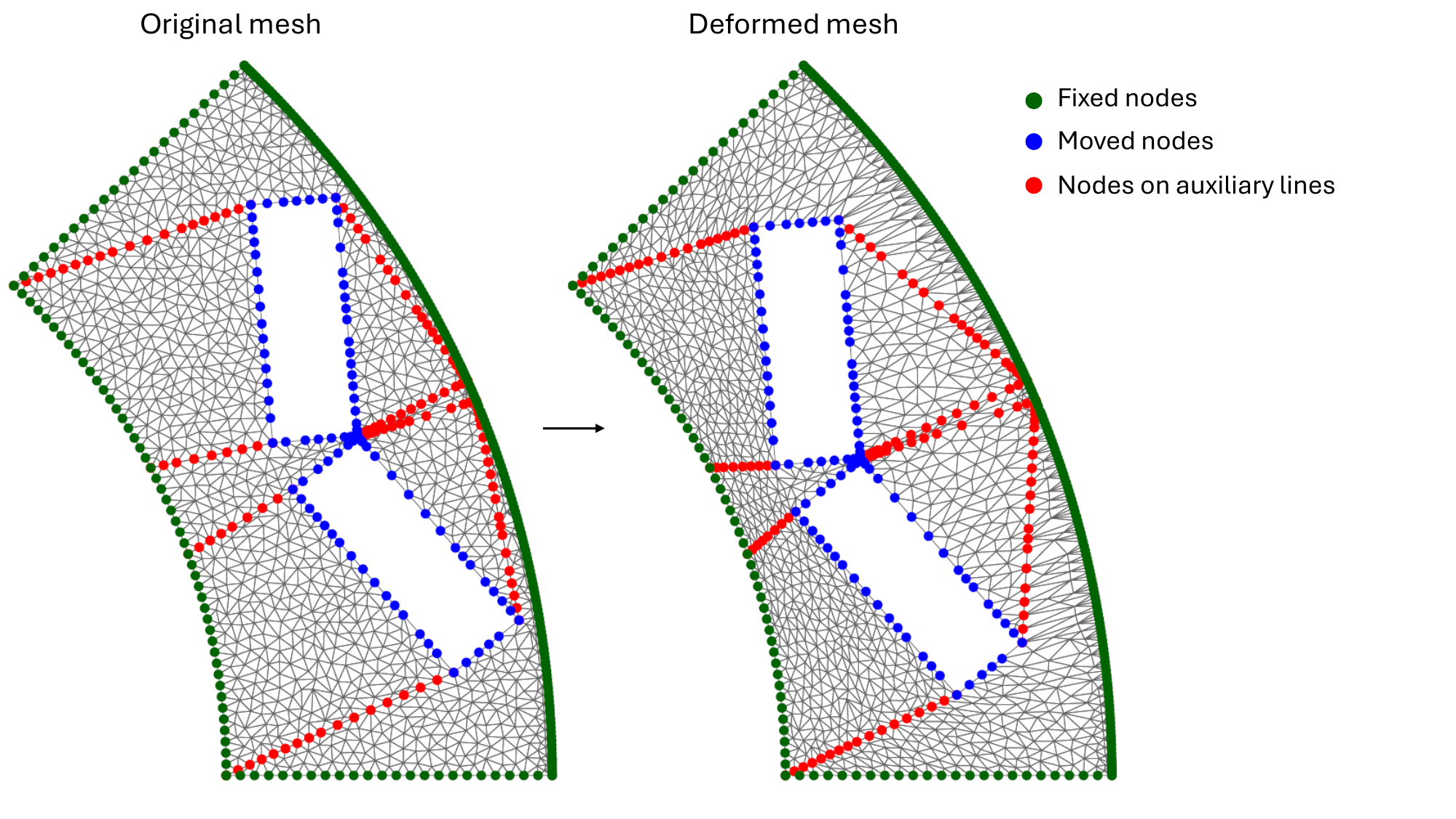}
	\caption{Original and deformed mesh using the Laplace-based method with auxiliary lines for magnet displacement} % links, unten, rechts, oben % trim = 130 40 50 20, clip,
	\label{fig:Prius2004_verschoben_Tiefer}
\end{figure}

An example solution is shown in Fig.~\ref{fig:Prius2004_verschoben_Tiefer}. The magnets are displaced toward the inner radius, the total number of elements remains unchanged, and the mesh can be overlaid with the binary method without remeshing.
%The finite-element-method is used to solve these equations. Nodes that should not be deformed (marked green) are integrated as dirichlet boundary conditions. The same is done for alle nodes on the magnets that are moved with a constant displacement (marked blue). For great movements of the magnet inside the rotor geometry, some triangular elements overlap. This is not suitable for the material allocation. Helper lines (marked red) are introduced to overcome this problem. They start on a fixed node and end on a edge node of a magnet. The solution for an example geometry is shown in figure \ref{fig:Prius2004_verschoben_Tiefer}. The magnets are moved towards the inner radius. Elements above the magnets are stretched and those under the magnets are compressed. The overall number of elements does not change, no elements overlap and the relative positioning stays the same. Therefore, a parametric optimization of the magnets can be overlayed with the binary method to optimize the remaining area. 

% -------------------------------------------------------------
% -------------------------------------------------------------
\subsection{Driving cycle reduction}

The driving cycle contains many operating points, so direct stochastic optimization would be computationally prohibitive. To mitigate this issue, a representative operating-point (ROP) approach is employed.

The k-means algorithm is used to determine the ROPs and their weightings~\cite{ray2000kmeans}. The number of ROPs equals the number of clusters $N_{\mathrm{C}}$, which must be specified in advance. K-means is selected because it provides compact centroid-based representatives with low computational overhead. The clustering minimizes the following objective function $J$:
%The driving cycle corresponds to a great number of operating points. A stochastic optimization over these is not suitable, because the high number of evaluated geometries leads to excessive computation time. An approach to reduce the number of points is considered. The k-means algorithm is used to calculate representative operating points (ROP) with specific weightings. The number of ROP equals the number of cluster $N\mx{C}$ and must be defined beforehand. $N\mx{OP}$ operating points from the driving cycle are used with their torque-speed-values $\vv{x}$. The cluster centers are called $\vv{s}$. The cluster definition is based on minimization of the objective function $J$:
\begin{align}
	%\mathrm{min} \; J = \sum\limits_{i=1}^{N\mx{C}} \sum\limits_{k=1}^{N\mx{OP}} || \vv{x}_k - \vv{s}_i ||^2
	J = \sum\limits_{i=1}^{N\mx{C}} \sum\limits_{\vv{x}_k \in \mathcal{C}_i} || \vv{x}_k - \vv{s}_i ||^2 \overset{!}{\rightarrow} \text{min}
\end{align}

To determine a suitable number of clusters, the inter-cluster distance $d_{\mathrm{inter}}$ is maximized so that the representative operating points are well distributed across the operating range.
%To find a suitable value for the number of clusters, comparison criteria for this stochastic approach are considered. The first criteria $d\mx{inter}$ is defined as the distance between the cluster centers (see eq. \ref{eq:d_inter}). This value should be maximized. 
\begin{align}
	d\mx{inter} = \min_{i \neq j}( |\vv{s}_i - \vv{s}_j|^2 ) \label{eq:d_inter}
\end{align}
The second criterion is the distance between all points inside a cluster and their corresponding center:
\begin{align}
	d\mx{intra} = \frac{1}{N\mx{C}} \sum\limits_{i=1}^{N\mx{C}} \sum\limits_{\vv{x}_k \in \mathcal{C}_i} | \vv{x}_k - \vv{s}_i  |^2 ,
\end{align}
where $\mathcal{C}_i$ denotes the set of operating points assigned to cluster $i$.
Minimization of this value identifies the optimal clustering. For further analysis, the ratio $d_{\mathrm{intra}} / d_{\mathrm{inter}}$ is used as a combined metric and should be minimized.

\begin{figure}[!t]
	\centering
	\includegraphics[width=0.48\textwidth]{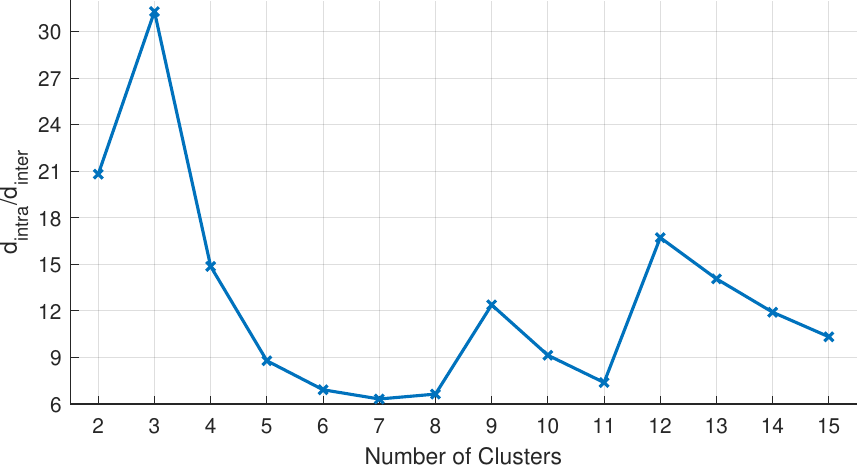}
	\caption{k-means clustering criteria ratio over different numbers of clusters for the WMTC driving cycle} % links, unten, rechts, oben % trim = 130 40 50 20, clip,
	\label{fig:k_means_intra_inter}
\end{figure}

\begin{figure}[!t]
	\centering
	\includegraphics[width=0.48\textwidth]{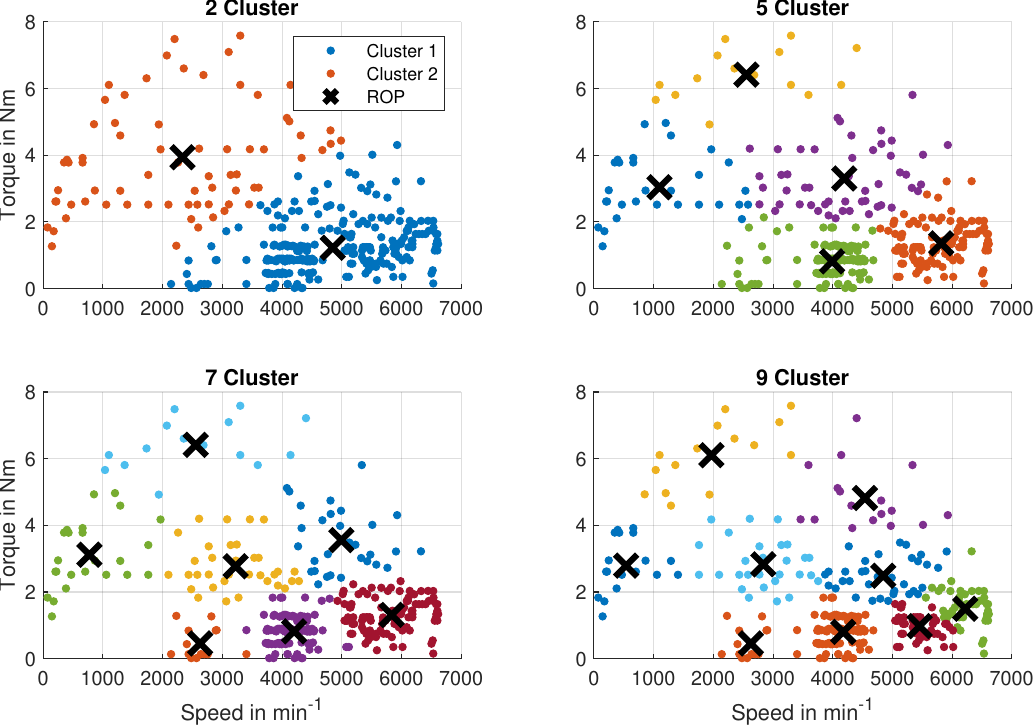}
	\caption{Motor operating points and representative operating points (ROPs) for different numbers of clusters} % links, unten, rechts, oben % trim = 130 40 50 20, clip,
	\label{fig:k_means_Cluster_vgl}
\end{figure}

A variation in the number of clusters is performed to evaluate the values of the defined ratio, as shown in Fig.~\ref{fig:k_means_intra_inter}. The resulting cluster assignments for four selected cluster numbers are illustrated in Fig.~\ref{fig:k_means_Cluster_vgl}.
%A minimization of this value is needed. For further analysis, the fraction $d\mx{intra} / d\mx{inter} \rightarrow \text{min}$ is used as a combination of both criteria. The motoric operating points of the given driving cycle are considered. A variation of the number of clusters is conducted to compare the values of the defined fraction (see fig. \ref{fig:k_means_intra_inter}). The clustering for the four marked values are shown in figure \ref{fig:k_means_Cluster_vgl}.

Five clusters are used for all subsequent calculations as a compromise between computational cost and accuracy. Although seven clusters would be preferable based on the combined criterion, this would increase computational effort by approximately \SI{40}{\%}.

The weightings $w$ for the five representative operating points are calculated from the electrical power $P_{\mathrm{el}}$ of all operating points within each cluster.
%Five cluster are used for all further calculations as a trade-off between computing time and accuracy. Seven cluster would be better based on the used fraction criteria, but this would be a \SI{40}{\%} rise of the computational costs. 

%The weightings $w$ for the five used ROP are calculated based on the electrical power $P\mx{el}$ of all operating points inside every cluster. It is defined as:
\begin{align}
	w_i = \frac{ \sum\limits_{j=1}^{N\mx{C,OP}} P_{\text{el,C,}ij} }{ \sum\limits_{k=1}^{N\mx{OP}}  P_{\text{el,}k} }.
\end{align}

The reference machine efficiency over the full driving cycle is \SI{92.62}{\%}. Using the ROPs and their weightings, the calculated efficiency is \SI{92.78}{\%}, a deviation of 0.16 percentage points. Higher cluster counts provide only marginal accuracy gains relative to the additional computation.

To validate the final design decision, the full driving-cycle efficiency of Rotor~8 was evaluated over all operating points after optimization. The result is \SI{92.54}{\%}, compared to \SI{92.62}{\%} for the reference machine, indicating magnet-material reduction at nearly unchanged full-cycle efficiency.
%The efficiency over all operating points on the driving cycle for the reference machine is \SI{92.6}{\%}. Using the ROP and their weighting, the calculated efficiency is \SI{92.8}{\%}.

% Rep: 92.8 %
% Full: 92.6 %

\section{Results}

All optimizations are performed using a genetic algorithm with a population size of 200 individuals.
The finite-element model employs approximately 15,000 triangular elements, resulting in a discretization error below \SI{0.5}{\%} for torque and core loss calculations.
While the observed torque improvements of \SIrange{0.02}{0.13}{\%} are within this numerical uncertainty, the consistent reduction in core losses (\SIrange{0.7}{1.5}{\%}) is above the reported mesh-discretization level.
The presented designs correspond to the best feasible candidates obtained under identical computational budgets and common stopping criteria across methods.
Accordingly, the primary performance indicator in this section is constraint-compliant magnet and loss reduction under cycle-based operation, with torque improvements reported as secondary effects.
%\lipsum[10]

% -------------------------------------------------------------
% -------------------------------------------------------------
\subsection{Optimization with fixed magnet geometry}

The first rotor optimization is performed with a fixed magnet geometry.
Three methods are employed to optimize the air-pocket layout: the binary (on/off) method, Normalized Gaussian Networks (NGnet), and a shape optimization approach using parameterized splines (SO).
Only the SO method uses the reference geometry as the initial design; all other methods start with a rotor containing no air pockets.
A genetic algorithm serves as the stochastic optimization core for all methods, using identical population sizes and stopping criteria.
Therefore, the comparison represents a controlled-budget benchmark aimed at practical method selection under fixed engineering resources.

\begin{figure}[!t]
	\centering
	\includegraphics[width=0.48\textwidth]{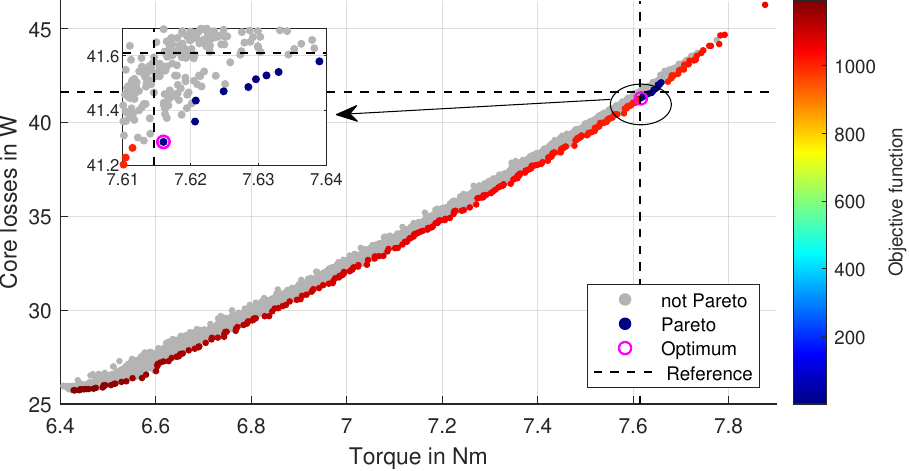}
	\caption{Results of all models from an optimization run using the binary method (Rotor~1), including color coding of the non-dominated set} % links, unten, rechts, oben % trim = 130 40 50 20, clip,
	\label{fig:Pareto_Binary_MagFest}
\end{figure}

\begin{figure}[!t]
	\centering
	\includegraphics[trim = 0 220 390 0,  clip, width=0.48\textwidth]{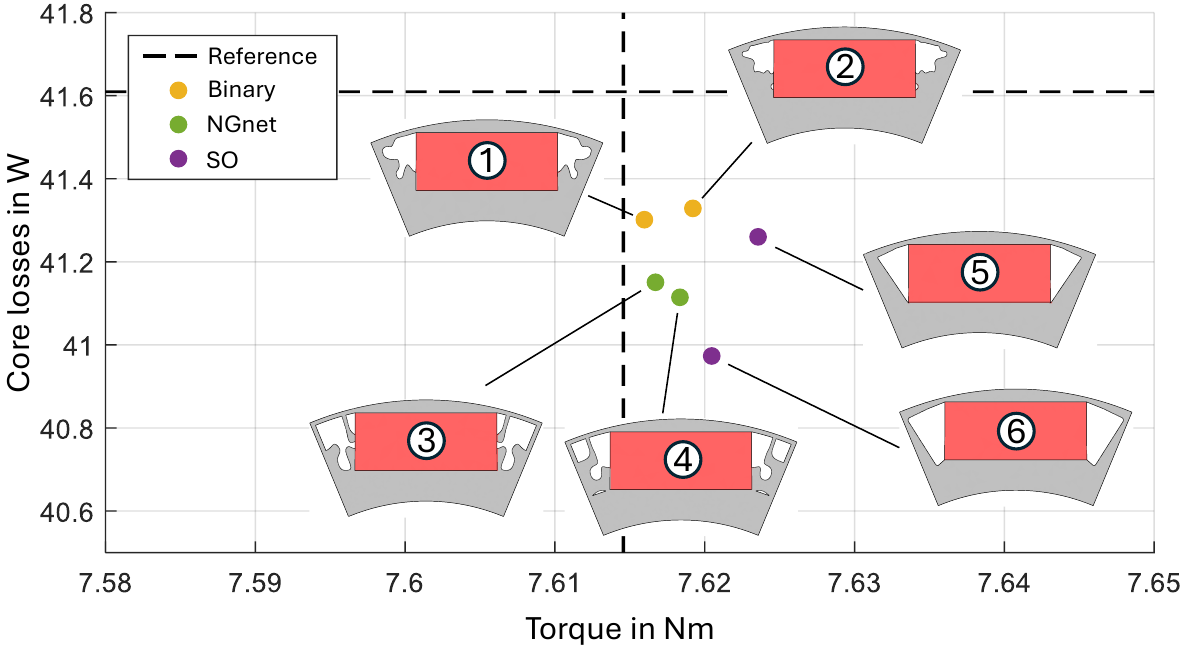}
	\caption{Results of shape and topology optimization over the driving cycle with fixed magnet geometry} % links, unten, rechts, oben % trim = 130 40 50 20, clip,
	\label{fig:Vgl_Verfahren_Magfest}
\end{figure}

Fig.~\ref{fig:Pareto_Binary_MagFest} shows the core losses (weighted over the representative operating points) versus the mean torque (at the operating point with the highest torque) for all geometries evaluated using the binary method.
The color map represents the value of the objective function.
Gray points are not part of the non-dominated set in the torque-loss plane.
The dotted lines indicate the corresponding values of the reference machine.
Geometries with $Z \ge 1000$ (red) violate at least one constraint.
Designs to the left of the dotted line exhibit insufficient torque, while red points in the top-right corner exceed the voltage limit.
Blue-colored geometries satisfy all constraints, and the purple-circled point achieves the best objective function value.
This design is labeled as Rotor~1 and is used for further comparison in Fig.~\ref{fig:Vgl_Verfahren_Magfest}.
%The first rotor optimization is conducted with a fixed magnet geometry. The binary method (on/off-method), Normalized Gaussian networks (NGnet) and a shape optimization with parametrized splines (SO) are used to optimize the airpockets. Only SO uses the reference geometry as a starting point, all other methods start without any airpockets inside the rotor. A genetic algorithm is used for all methods as the stochastic core. Figure \ref{fig:Pareto_Binary_MagFest} shows core losses (weighted over the ROP) and mean torque (operating point with highest torque) for all simulated geometries using the binary method. The coloring indicates the value of the objective function. All grey points are not part of the pareto front. The dotted lines mark the values from the reference machine. Geometries with $Z >= 1000$ (red) violate at least one constraint. The geometries left from the dotted line achieve a too low torque. All red points in the top right corner exceed the voltage limit. Blue colored geometries are suitable and the purple circled point has the best objective value. This result is labeled as rotor 1 and used for further comparison in figure \ref{fig:Vgl_Verfahren_Magfest}.

The binary method is applied again with a different set of hyperparameters to obtain Rotor~2.
The use of NGnet results in Rotor~3 and Rotor~4 geometries, while the shape optimization (SO) approach produces Rotor~5 and Rotor~6 topologies.
Overall, all optimized geometries exhibit higher mean torque and lower core losses compared to the reference machine.
Among them, Rotor~5 achieves the highest torque, whereas Rotor~6 has the lowest core losses.
However, the overall variation in performance across all optimized geometries is relatively small, indicating that the different optimization methods converge to similar design regions in the considered design space.
%The binary method is used again with a different set of hyper parameters to find rotor 2. Usage of NGnet leads to the rotor geometries 3 and 4. The optimization with SO yields the rotor topologies 5 and 6. Overall, all geometries achieve higher mean torque and lower losses than the reference machine. Rotor 5 has the highest torque and rotor 6 the lowest core losses. But the range of the performance values is really low over all found geometries.

% -------------------------------------------------------------
% -------------------------------------------------------------
\subsection{Optimization with variable magnet geometry}

\begin{figure}[!t]
	\centering
	\includegraphics[trim = 0 220 380 0,  clip, width=0.48\textwidth]{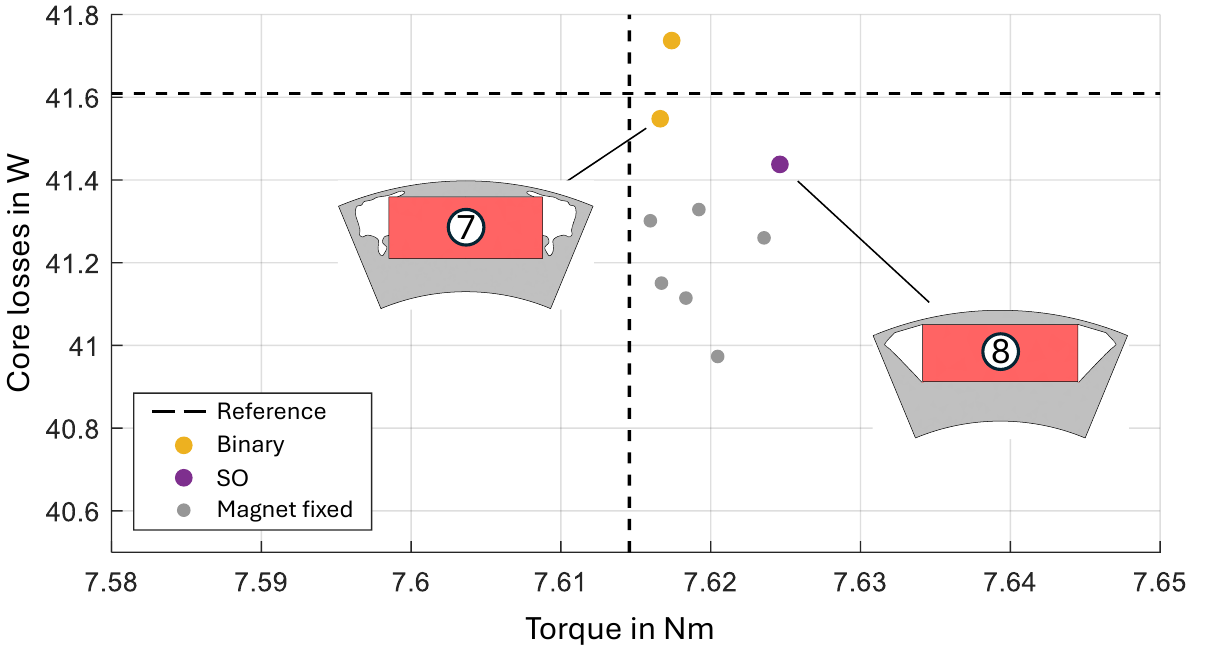}
	\caption{Results of shape and topology optimization over the driving cycle with variable magnet geometry} % links, unten, rechts, oben % trim = 130 40 50 20, clip,
	\label{fig:Vgl_Verfahren_Magvar}
\end{figure}

A parametric optimization of the magnet geometry is combined with the topology optimization of the air pockets.
For the binary method, the mesh deformation strategy described in Section~\ref{ch:Mesh_deform} is employed, whereas it cannot be applied in conjunction with NGnet.
The shape optimization (SO) approach allows simultaneous optimization of both the magnet parameters and the air-pocket splines within a single optimization loop.
The results are illustrated in Fig.~\ref{fig:Vgl_Verfahren_Magvar}, where the gray points correspond to the results of the previous optimization performed with fixed magnets.
%A parametric optimization of the magnet geometry is combined with the topology optimization of the airpockets. The mesh deformation from chapter \ref{ch:Mesh_deform} is used for the binary method, but cannot be integrated with NGnet. SO can be used directly with parameters for the magnet and for the airpocket splines. The results are visualized in figure \ref{fig:Vgl_Verfahren_Magvar}. The grey points are the results from the prior optimization with fixed magnets. 

The variable-magnet study yields feasible designs with reduced magnet cross-sectional area compared to the reference machine while still meeting the required torque.
However, one geometry obtained using the binary method exhibits increased core losses.
In contrast, the shape-optimized geometry not only exceeds the torque requirement but also achieves lower core losses and the smallest magnet volume.
%All three results have a smaller magnet cross-section area compared to the reference and exceed the required torque. However, one geometry of the binary method leads to increased core losses. The shape-optimized geometry exceeds the torque requirement, has lower core losses, and the smallest magnets.

\begin{table}[!t]
	% increase table row spacing, adjust to taste
	\renewcommand{\arraystretch}{1.3}
	\caption{Relative changes of torque, losses, and magnet volume after optimization compared to the reference machine}
	\label{tab:Results_table}
	\centering
	\begin{tabular}{|c|c|c|c|c|c|}
		\hline
		Rotor & Method & Variables & Torque         & Losses         & Magnet Vol.   \\
		\hline
		1     & Binary & 405       & +\SI{0.02}{\%} & \SI{-0.74}{\%} & --            \\
		2     & Binary & 259       & +\SI{0.06}{\%} & \SI{-0.67}{\%} & --            \\
		\hline
		3     & NGnet  & 32        & +\SI{0.03}{\%} & \SI{-1.1}{\%}  & --            \\
		4     & NGnet  & 32        & +\SI{0.05}{\%} & \SI{-1.2}{\%}  & --            \\
		\hline
		5     & SO     & 7         & +\SI{0.12}{\%} & \SI{-0.8}{\%}  & --            \\
		6     & SO     & 5         & +\SI{0.08}{\%} & \SI{-1.5}{\%}  & --            \\
		\hline
		\hline
		7     & Binary & 346       & +\SI{0.03}{\%} & \SI{-0.15}{\%} & \SI{-3.6}{\%} \\
		\hline
		8     & SO     & 7         & +\SI{0.13}{\%} & \SI{-0.4}{\%}  & \SI{-10}{\%}  \\
		\hline
	\end{tabular}
\end{table}

From an industrial perspective, the key result is that permanent-magnet demand can be reduced by up to \SI{10}{\%} while maintaining feasible operation with respect to torque, inverter-voltage, and mechanical-strength limits.

% -------------------------------------------------------------
% -------------------------------------------------------------
\subsection{Mechanical strength analysis}

\begin{figure}[!t]
	\centering
	\includegraphics[width=0.4\textwidth]{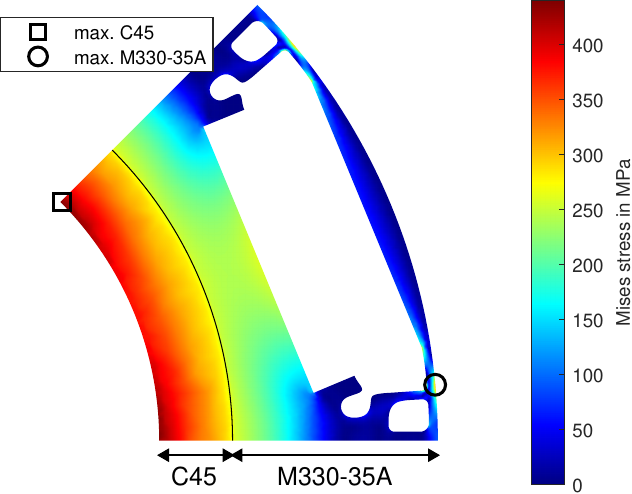}
	\caption{Von Mises stress distribution of Rotor~4 at maximum overspeed} % links, unten, rechts, oben % trim = 130 40 50 20, clip,
	\label{fig:Mises_NGnet_final}
\end{figure}

For all candidate designs, a two-dimensional plane-stress mechanical finite-element analysis is performed at the maximum operating speed increased by \SI{20}{\%}.
The von Mises stress distribution is evaluated in the rotor core, the magnets, and the shaft.
Fig.~\ref{fig:Mises_NGnet_final} illustrates the von Mises stress distribution of Rotor~4 obtained with the NGnet-based topology optimization.

For all optimized geometries, the maximum von Mises stress remains below the corresponding \SI{0.2}{\%} proof strength of the electrical steel and shaft materials defined in Section~II, such that the mechanical constraint term $Z_{\mathrm{mech}}$ remains zero.
The air-pocket smoothing and block-filtering strategies described in Section~III are essential to avoid sharp notches and local stress peaks.
The reported stress values are based on a two-dimensional plane-stress model; additional three-dimensional features such as end-region stress concentration are not covered in this study.

% -------------------------------------------------------------
% -------------------------------------------------------------
\subsection{Discussion}

The numerical results show that the proposed framework can systematically find feasible designs with reduced weighted core losses and/or reduced magnet volume compared to the reference design.
For fixed magnets, all three optimization methods (binary, NGnet, and SO) yield rotor geometries with slightly increased torque and reduced core losses at the selected representative operating points, with NGnet and SO providing the best trade-off between performance and number of design variables.
The binary method uses 259-405 design variables (mesh element densities), NGnet employs 32 continuous variables, while SO uses only 5-7 parametric spline variables.
This strong reduction in dimensionality explains the faster convergence behavior of SO under the same optimization budget.

When the magnet geometry is included as an additional degree of freedom, the achievable reduction in magnet volume is up to \SI{10}{\%} without violating torque, voltage, or mechanical constraints.
For Rotor~8, the full-cycle efficiency is slightly lower than the reference by 0.08 percentage points (see Section III), indicating a near-efficiency-preserving magnet-material reduction rather than a strict full-cycle efficiency gain.
In particular, the shape optimization approach proves to be the most effective compromise between design flexibility and computational effort, as summarized in Table~\ref{tab:Results_table}.
While the binary topology optimization offers a significantly larger design space, the high dimensionality of the problem (over 400 design variables) poses a challenge for the genetic algorithm to explore effectively within the given population and generation limits.
In contrast, the shape optimization approach benefits from a much lower number of variables and starts from a well-defined initial reference geometry, allowing efficient convergence to designs that closely align with the physical flux paths.
The k-means-based driving-cycle reduction enables these studies with acceptable computation time while preserving good approximation accuracy.
Overall, the strongest outcome is not a large torque increase, but a robust constraint-compliant magnet reduction supported by experimental validation.
The proposed pipeline is intended for early-to-mid design stages, where rapid narrowing to physically feasible candidates is critical before more expensive three-dimensional and thermal co-optimization steps.

\section{Experimental validation}

\begin{figure}[!t]
	\centering
	\subfloat[Rotor 4]{\includegraphics[ width=0.2\textwidth]{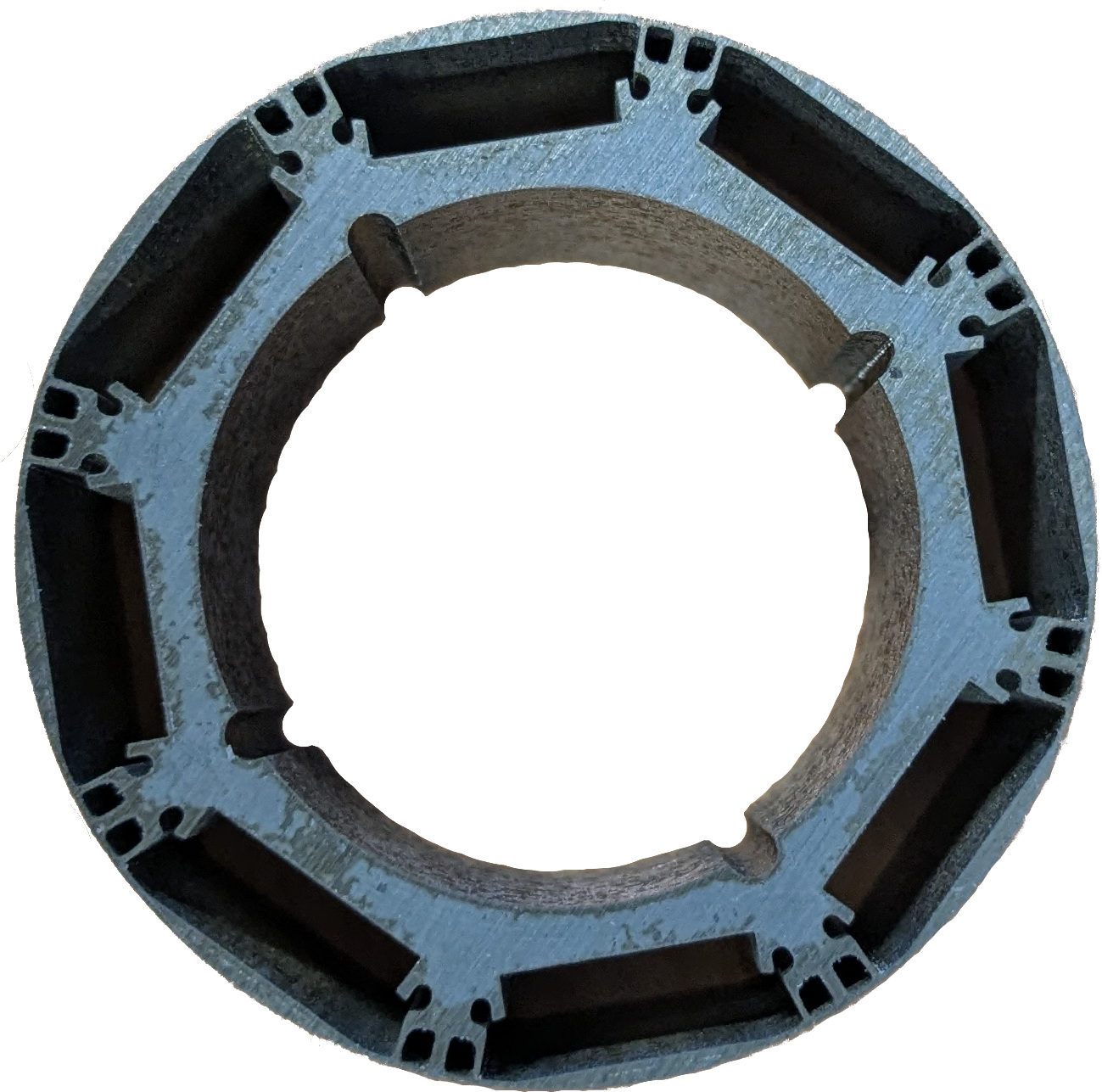}%
		\label{Rotor_4_manuf}}
	\hfil
	\subfloat[Rotor 8]{\includegraphics[ width=0.2\textwidth]{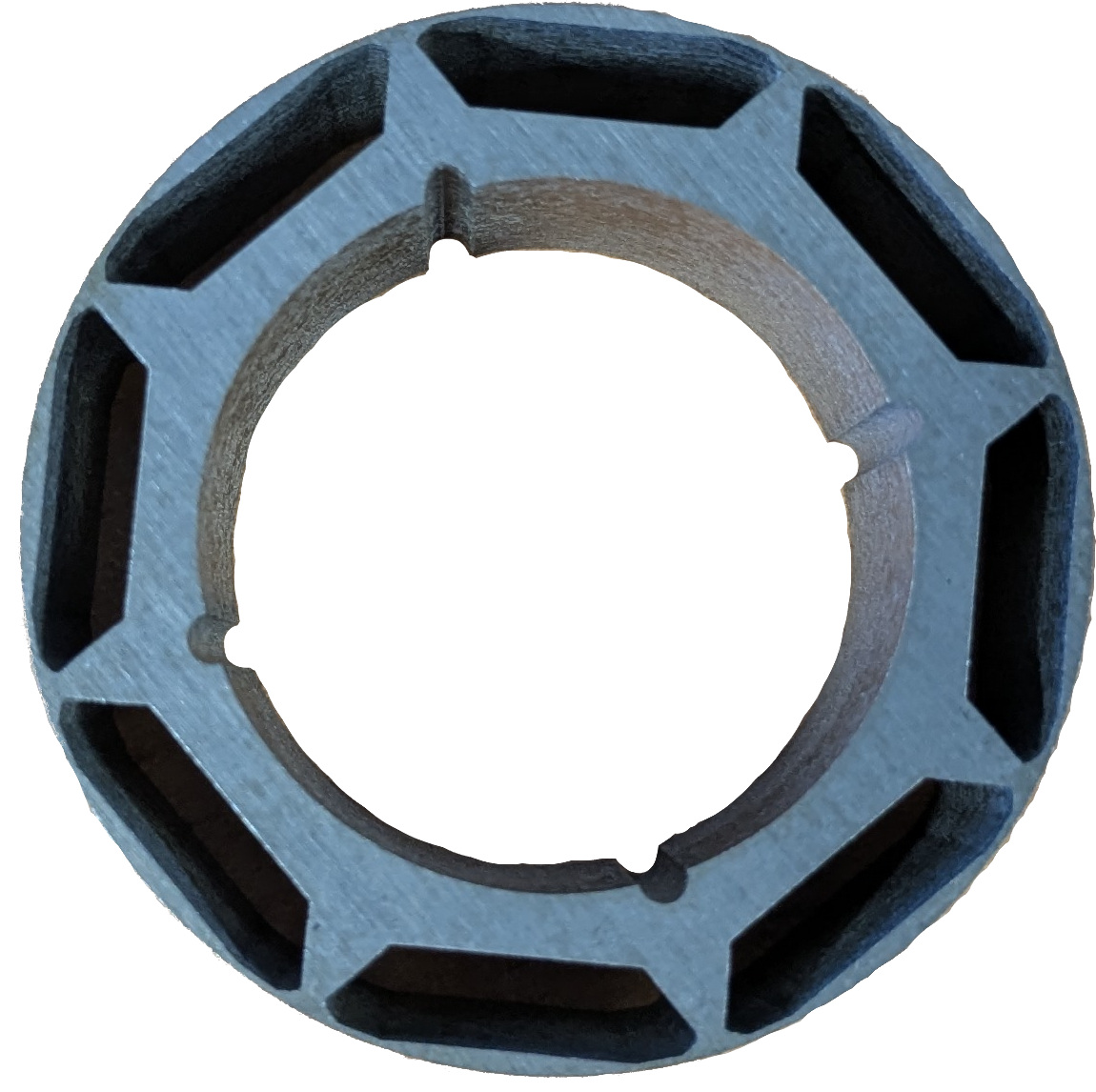}%
		\label{Rotor_8_manuf}}

	\caption{Manufactured rotor cores (Rotor~4 and Rotor~8) before magnet insertion}
	\label{fig:Manufactered_rotors}
\end{figure}

For experimental validation, rotor topologies 4 and 8 are manufactured (Fig.~\ref{fig:Manufactered_rotors}).
Rotor~4 validates the NGnet-based branch with fixed magnets, while Rotor~8 represents the shape-optimization branch with variable magnet geometry.

\begin{figure}[!t]
	\centering
	\includegraphics[trim =  0 170 230 0,  clip, clip,width=0.48\textwidth]{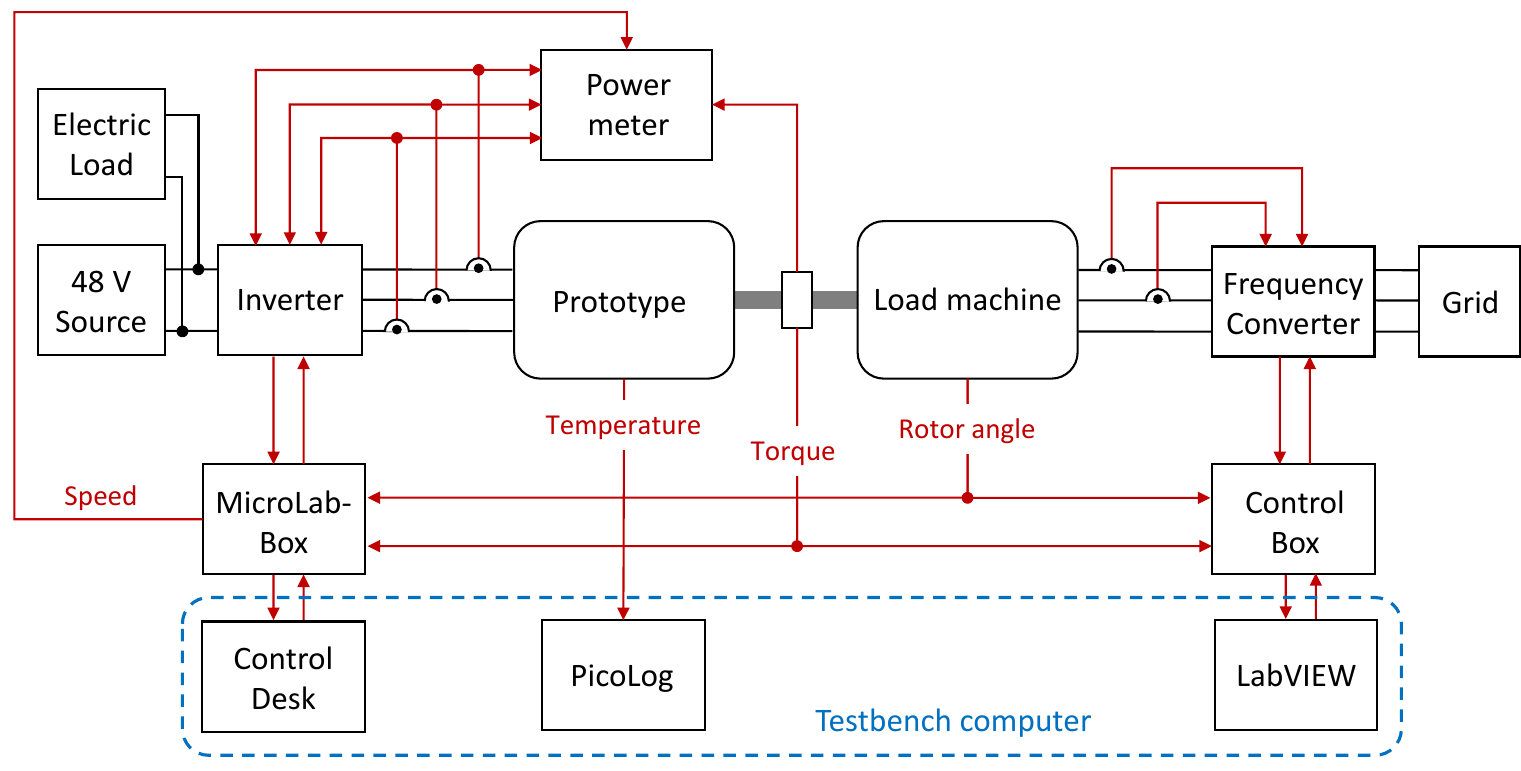} % links, unten, rechts, oben
	\caption{Schematic diagram of the test bench used for validation}
\label{Aufbau_Pruefstand}
\end{figure}

\begin{figure}[!t]
	\centering
	\includegraphics[trim =  0 280 170 0,  clip, clip,width=0.48\textwidth]{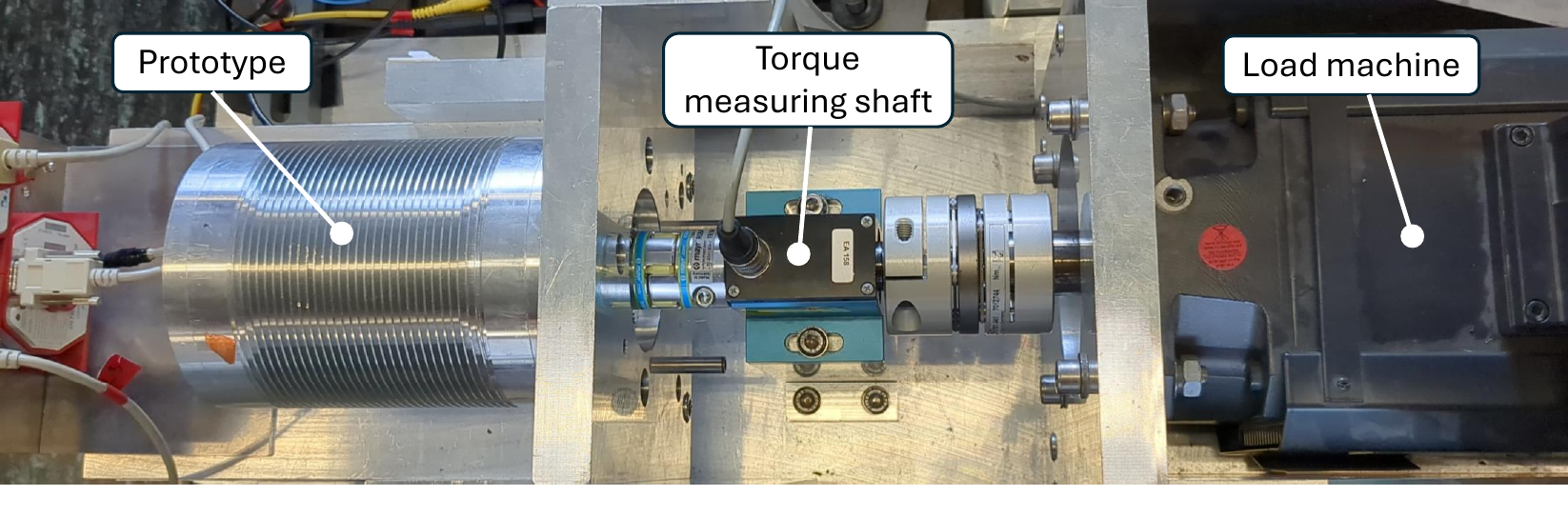} % links, unten, rechts, oben
	\caption{Coupled traction and load machines on the test bench}
\label{Pruefstand_real}
\end{figure}

An experimental test bench is built for the measurements and is visualized in Fig.~\ref{Aufbau_Pruefstand}.
Phase voltages and currents, electrical power, torque, rotor angle, speed, and stator-winding temperature are measured.
A second synchronous machine is used as the load machine, as shown in Fig.~\ref{Pruefstand_real}.

The experimental section focuses on simulation-versus-measurement consistency for Rotors~4 and 8;
relative improvements to the reference design are assessed by simulation under the same evaluation pipeline.

% -------------------------------------------------------------
% -------------------------------------------------------------
\subsection{Back-EMF measurement}

\begin{figure}[!t]
	\centering
	\subfloat[Full period]{\includegraphics[ width=0.33\textwidth]{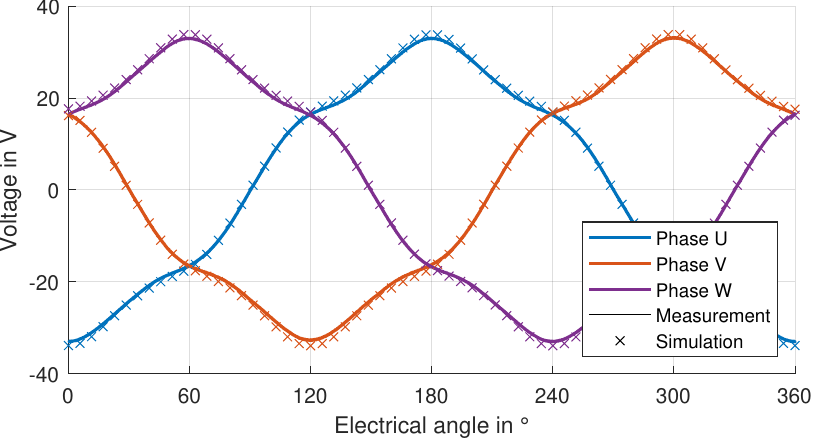}%
		\label{Full}}
	\hfil
	\subfloat[Zoom]{\includegraphics[ width=0.14\textwidth]{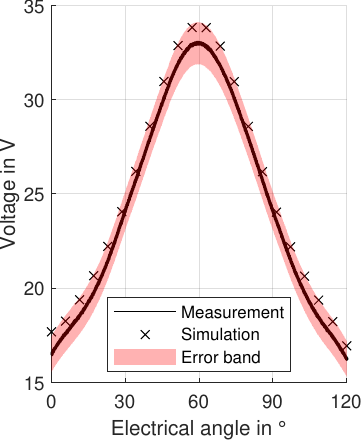}%
		\label{Zoom}}

	\caption{Measured and simulated back-EMF at \SI{22}{^\circ C} and $\SI{3000}{min^{-1}}$ for Rotor~4, including measurement uncertainty band}
	\label{fig:Back_emf_meas_rotor_NGnet}
\end{figure}

\begin{figure}[!t]
	\centering
	\subfloat[Full period]{\includegraphics[ width=0.33\textwidth]{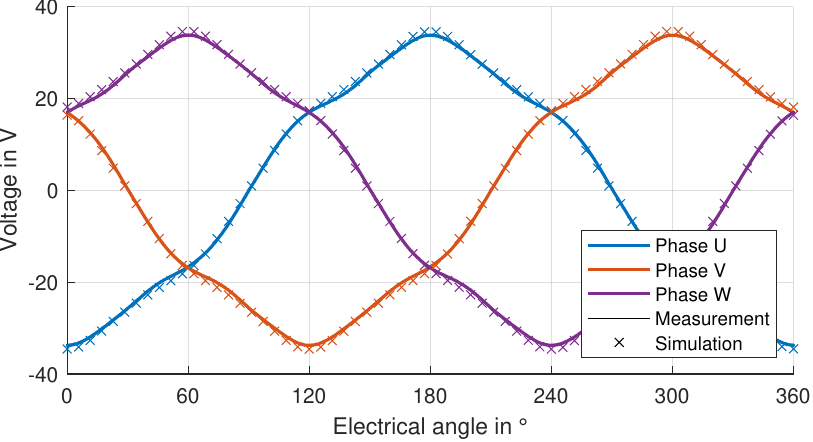}%
		\label{Full_rotor8}}
	\hfil
	\subfloat[Zoom]{\includegraphics[ width=0.14\textwidth]{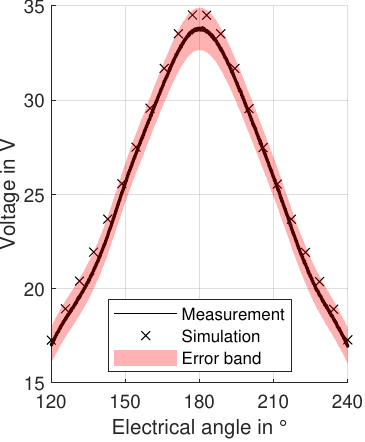}%
		\label{Zoom_rotor8}}

	\caption{Measured and simulated back-EMF at \SI{22}{^\circ C} and $\SI{3000}{min^{-1}}$ for Rotor~8, including measurement uncertainty band}
	\label{fig:Back_emf_meas_rotor_FO}
\end{figure}

A no-load experiment is conducted to measure the back-EMF waveforms with a Rohde and Schwarz RTM3004 oscilloscope ($\pm \SI{1.5}{\%}$) and Testec differential probes ($\pm \SI{1}{\%}$).
The measurements are shown in Fig.~\ref{fig:Back_emf_meas_rotor_NGnet} and Fig.~\ref{fig:Back_emf_meas_rotor_FO} together with the simulation results, and all simulated points lie within the error band.

% -------------------------------------------------------------
% -------------------------------------------------------------
\subsection{Torque measurement}

Torque measurements are carried out under steady-state conditions for several operating points covering the main region of the efficiency map.
The electromagnetic torque is obtained from a calibrated torque transducer, while the simulated torque values are interpolated to the same operating points.

The relative deviation between measurement and simulation remains below \SI{3}{\%} for all investigated operating points. The largest deviations occur at low torque and low speed, where inverter nonidealities and friction effects are more prominent.

% -------------------------------------------------------------
% -------------------------------------------------------------
\subsection{Efficiency map measurement}

\begin{figure}[!t]
	\centering
	\subfloat[Simulation]{\includegraphics[trim = 0 0 0 0,  clip, width=0.23\textwidth]{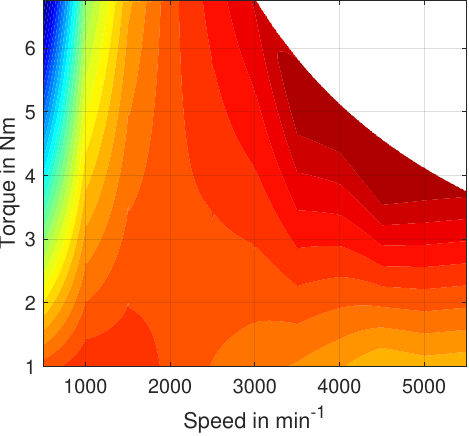}%
		\label{Rotor_4_eff_diff}}
	\hfil
	\subfloat[Measurement]{\includegraphics[trim = 5 0 0 0,  clip, width=0.25\textwidth]{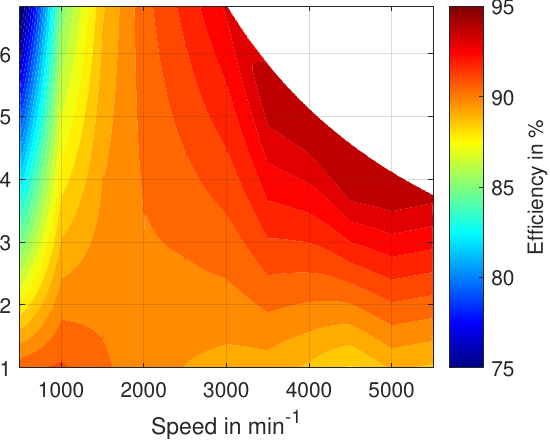}%
		\label{Rotor_8_eff_diff}}

	\caption{Efficiency maps from simulation and measurement for Rotor~8}
	\label{fig:Rotor_FO_eta_measure}
\end{figure}

\begin{figure}[!t]
	\centering
	\subfloat[Rotor 4]{\includegraphics[trim = 0 0 55 0,  clip, width=0.205\textwidth]{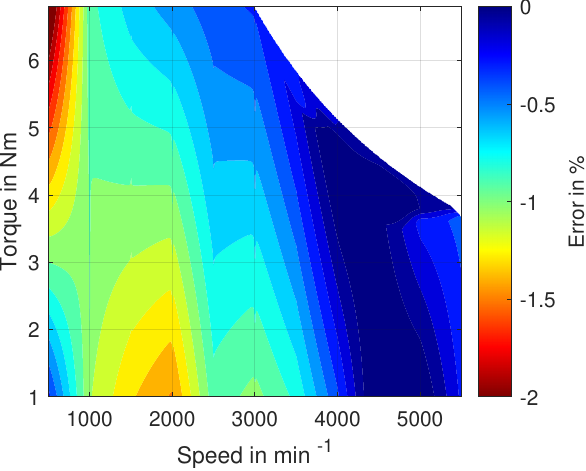}%
		\label{Rotor_4_eta_diff}}
	\hfil
	\subfloat[Rotor 8]{\includegraphics[trim = 20 0 0 0,  clip, width=0.24\textwidth]{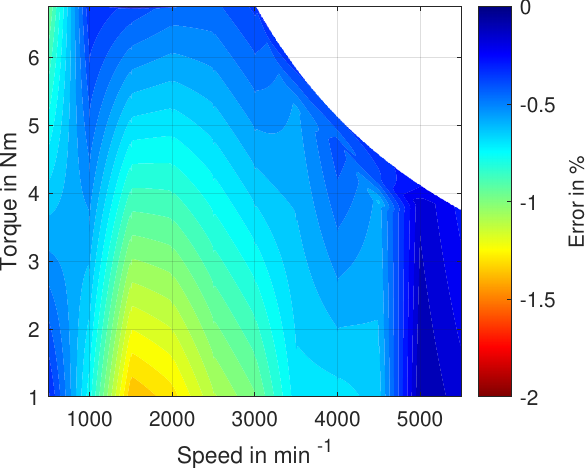}%
		\label{Rotor_8_eta_diff}}

	\caption{Efficiency map differences between simulation and measurement for both rotors}
	\label{fig:Efficiency_difference}
\end{figure}

The efficiency maps of Rotor~4 and Rotor~8 are obtained from steady-state measurements by sweeping torque and speed over the relevant operating range.
Figure \ref{fig:Rotor_FO_eta_measure} shows the efficiency maps for Rotor~8 from simulation and measurement, while Fig.~\ref{fig:Efficiency_difference} shows the deviations for both rotors.
Rotor~4 has a maximum deviation of \SI{-2}{\%}, and Rotor~8 exhibits a maximum deviation of \SI{-1.3}{\%}.
In both cases, the simulated efficiency exceeds the measured values.

\begin{figure}[!t]
	\centering
	\includegraphics[trim =  0 0 0 0,  clip, clip,width=0.39\textwidth]{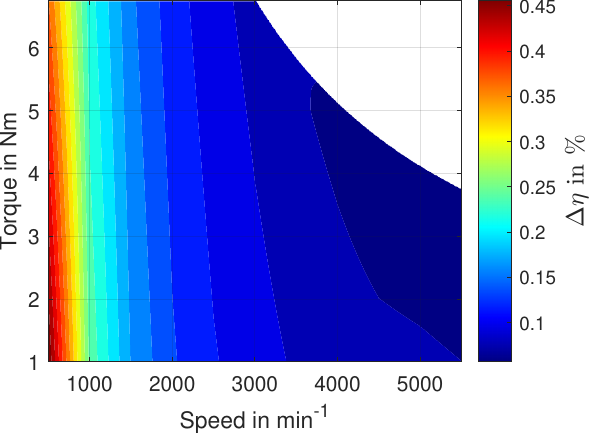} % links, unten, rechts, oben
	\caption{Uncertainty of the efficiency map measurement for the used test bench}
	\label{eta_map_tolerance}
\end{figure}

The electrical power $P$ was determined using a Zimmer LMG670 precision power analyzer.
To determine the total uncertainty of the power measurement, the Gaussian law of error propagation was applied~\cite{jcgm2008}.
The combined uncertainty $\Delta P$ is calculated as:
\begin{equation}
	\Delta P = \sqrt{ \left( \frac{\partial P}{\partial U} \cdot \Delta U \right)^2 + \left( \frac{\partial P}{\partial I} \cdot \Delta I \right)^2 }
\end{equation}

where $\Delta U$ and $\Delta I$ represent the individual accuracy specifications provided by the manufacturer (\SI{0.015}{\%} of the reading + \SI{0.03}{\%} of the range).
The same calculation is used for mechanical power and efficiency.
Figure \ref{eta_map_tolerance} shows the total uncertainty for the efficiency measurement over the full map, with a maximum of \SI{0.45}{\%} at low speeds.

%Copper losses are calculated from phase currents and winding resistance, while iron and mechanical losses are determined from the difference between electrical input power and mechanical output power at the shaft.

%\begin{figure}[!t]
%	\centering
%	\includegraphics[width=0.45\textwidth]{figures/Deutsch/eta_Kennfeld_Vgl_FO.pdf}
%	\caption{Reference machine for optimization} % links, unten, rechts, oben % trim = 130 40 50 20, clip,
%	\label{fig:eta_Kennfeld_Vgl_FO}
%\end{figure}

%\begin{align}
%	\Delta f &= \sqrt{ \left( \frac{\partial f}{\partial x} \cdot \Delta x \right)^2 
%		+  \left( \frac{\partial f}{\partial y} \cdot \Delta y \right)^2 }
%\end{align}

% -------------------------------------------------------------
% -------------------------------------------------------------
\subsection{Discussion}
\label{sec:meas_discussion}

The discrepancies between the experimental measurements and the simulation results can be attributed to the following factors:

\begin{itemize}
	\item Uncertainty in magnet temperature: The precise magnet temperature is not known, as temperature measurements are conducted at the stator windings.
	\item Material variances: Permanent magnets exhibit inherent tolerances and variances in their magnetic properties.
	\item Manufacturing influences: Fabrication processes can negatively affect core losses (e.g., through thermal stress).
	\item Inverter supply: Operation via a frequency inverter leads to higher core losses compared to sinusoidal excitation due to harmonic content.
	\item Neglected end effects: Three-dimensional end effects are neglected in the simulation.
	\item Measurement tolerances: Accuracy limits and tolerances of the measurement equipment may contribute to further deviations.
\end{itemize}

Most of these factors increase actual losses, which is consistent with the numerically determined efficiency values exceeding the measured values across all operating points.

\section{Conclusion}

This paper presented a shape and topology optimization framework for interior permanent magnet synchronous machines that accounts for driving-cycle performance and electromagnetic, mechanical, and inverter-related constraints.
The proposed approach combines a gradient-free genetic algorithm with binary, NGnet, and spline-based shape parameterizations.
A k-means clustering technique reduces the number of operating points while preserving efficiency-evaluation accuracy.

For the electric-scooter traction application, all three optimization methods yield rotor geometries with slightly increased torque and reduced weighted losses at the representative operating points compared with the reference machine.
When the magnet geometry is included, permanent-magnet volume is reduced by up to \SI{10}{\%} without violating torque, voltage, or mechanical-strength constraints.
The best variable-magnet candidate (Rotor~8) retains nearly unchanged full-cycle efficiency (\SI{92.54}{\%} versus \SI{92.62}{\%} for the reference), and the shape optimization approach provides the best compromise between performance, magnet reduction, and number of design variables.

Experimental back-EMF and efficiency-map measurements on two manufactured rotor prototypes confirm the electromagnetic models and support the proposed constraint-aware optimization pipeline.
The agreement between simulations and measurements is within a few percent over the relevant operating range, supporting the framework for practical design optimization and reliable design-ranking decisions under realistic driving-cycle conditions.

Future work will focus on three-dimensional models, additional loss components such as inverter induced core losses, cooling-system design, and extended sensitivity studies.

\ifCLASSOPTIONcaptionsoff
  \newpage
\fi

\end{document}